\documentclass[aps,prd,reprint,twocolumn,showpacs,superscriptaddress]{revtex4-1}  

\usepackage{epsfig}
\usepackage{enumitem}

\usepackage{verbatim}
\usepackage{graphics}     
\usepackage{graphicx}     
\usepackage{subfigure}
\usepackage{longtable}     
\usepackage{url}          
\usepackage{bm}           
\usepackage{textpos}
\usepackage{amsfonts,amsmath,amssymb,mathrsfs,bm,amsthm}
\usepackage{float}
\usepackage{booktabs}
\usepackage{array}
\usepackage{tabu}
\usepackage{dcolumn}
\usepackage{rotating}

\newcommand{\RN}[1]{%
  \textup{\uppercase\expandafter{\romannumeral#1}}%
}

\usepackage{nicefrac}
\usepackage{amsthm}
\usepackage{color}
\usepackage{cancel}
\usepackage{appendix}
\usepackage{makecell}
\usepackage{upgreek}

\usepackage[colorlinks=true,linkcolor=black,citecolor=blue,urlcolor=blue,bookmarksopen]{hyperref}

\newcommand{\appropto}{\mathrel{\vcenter{
  \offinterlineskip\halign{\hfil$##$\cr
    \propto\cr\noalign{\kern2pt}\sim\cr\noalign{\kern-2pt}}}}}
\renewcommand{\v}[1]{\boldsymbol{#1}}		

\def\be{\begin{equation}}
\def\ee{\end{equation}}
\def\ba{\begin{eqnarray}}
\def\ea{\end{eqnarray}}
\def\ge{\mathrel{\raise.3ex\hbox{$>$\kern-.75em\lower1ex\hbox{$\sim$}}}}
\def\la{\mathrel{\raise.3ex\hbox{$<$\kern-.75em\lower1ex\hbox{$\sim$}}}}

\def\simgt{\mathrel{\raise.3ex\hbox{$>$\kern-.75em\lower1ex\hbox{$\sim$}}}}
\def\simlt{\mathrel{\raise.3ex\hbox{$<$\kern-.75em\lower1ex\hbox{$\sim$}}}}

\newcommand{\bi}[1]{\bibitem{#1}}

\newcommand{\nc}{\newcommand}

\nc{\gone}{\bar g_{\pi NN}^{(1)}}
\nc{\gzero}{\bar g_{\pi NN}^{(0)}}
\nc{\al}{\alpha}
\nc{\ga}{\gamma}
\nc{\de}{\delta}
\nc{\ep}{\epsilon}
\nc{\ze}{\zeta}
\nc{\et}{\eta}
\nc{\ka}{\kappa}
\nc{\rh}{\rho}
\nc{\si}{\sigma}
\nc{\ta}{\tau}
\nc{\up}{\upsilon}
\nc{\ph}{\phi}
\nc{\ch}{\chi}
\nc{\ps}{\psi}
\nc{\om}{\omega}
\nc{\Ga}{\Gamma}
\nc{\De}{\Delta}
\nc{\La}{\Lambda}
\nc{\Si}{\Sigma}
\nc{\Up}{\Upsilon}
\nc{\Ph}{\Phi}
\nc{\Ps}{\Psi}
\nc{\Om}{\Omega}
\nc{\ptl}{\partial}
\nc{\del}{\nabla}
\nc{\ov}{\overline}
\nc{\newcaption}[1]{\centerline{\parbox{15cm}{\caption{#1}}}}
\nc{\us}{U(1)$_S$}
\nc{\Rg}{$R_{\gamma\gamma}$}
\nc{\bbnu}{\beta\beta_{0\nu}}

\def\beq{\begin{equation}}
\def\eeq{\end{equation}}
\def\bmat{\begin{displaymath}}
\def\emat{\end{displaymath}}
\def\bear{\begin{eqnarray}}
\def\eear{\end{eqnarray}}
\def\ba{\begin{eqnarray}}
\def\ea{\end{eqnarray}}
\def\bery{\begin{array}}
\def\ery{\end{array}}
\def\bit{\begin{itemize}}
\def\eit{\end{itemize}}
\def\ben{\begin{enumerate}}
\def\een{\end{enumerate}}
\def\btab{\begin{tabular}}
\def\etab{\end{tabular}}
\def\btbl{\begin{table}}
\def\etbl{\end{table}}
\def\bfig{\begin{figure}[htb]}
\def\efig{\end{figure}}
\def\bpic{\begin{picture}}
\def\epic{\end{picture}}

\def\ga{\mathrel{\raise.3ex\hbox{$>$\kern-.75em\lower1ex\hbox{$\sim$}}}}
\def\la{\mathrel{\raise.3ex\hbox{$<$\kern-.75em\lower1ex\hbox{$\sim$}}}}
\def\gappeq{\mathrel{\rlap {\raise.5ex\hbox{$>$}}
{\lower.5ex\hbox{$\sim$}}}}
\def\lappeq{\mathrel{\rlap{\raise.5ex\hbox{$<$}}
{\lower.5ex\hbox{$\sim$}}}}

\def\gyr{{\rm \, G\kern-0.125em yr}}
\def\mev{{\rm \, Me\kern-0.125em V}}
\def\gev{{\rm \, Ge\kern-0.125em V}}
\def\tev{{\rm \, Te\kern-0.125em V}}

\def\lsim{\mathrel{\rlap{\lower4pt\hbox{\hskip1pt$\sim$}}
    \raise1pt\hbox{$<$}}}                
\def\gsim{\mathrel{\rlap{\lower4pt\hbox{\hskip1pt$\sim$}}
    \raise1pt\hbox{$>$}}}                

\definecolor{jazzberryjam}{rgb}{0.65, 0.04, 0.37}        

\begin{document}

\title{Sensitivity of EDM experiments in paramagnetic atoms and molecules\\ to hadronic \boldmath{$CP$} violation}

\author{V.~V.~Flambaum}
\affiliation{School of Physics, University of New South Wales, Sydney 2052, Australia}

\author{M.~Pospelov}
\affiliation{School of Physics and Astronomy, University of Minnesota, Minneapolis, MN 55455, USA}
\affiliation{William I. Fine Theoretical Physics Institute, School of Physics and Astronomy, University of Minnesota, Minneapolis, MN 55455, USA}
\affiliation{Perimeter Institute for Theoretical Physics, Waterloo, ON N2J 2W9, Canada}
\author{A.~Ritz}
\affiliation{Department of Physics and Astronomy, University of Victoria, Victoria, BC V8P 5C2, Canada}

\author{Y.~V.~Stadnik}
\affiliation{School of Physics, University of New South Wales, Sydney 2052, Australia}
\affiliation{Helmholtz Institute Mainz, Johannes Gutenberg University, 55128 Mainz, Germany}
\affiliation{Kavli Institute for the Physics and Mathematics of the Universe (WPI), The University of Tokyo Institutes for Advanced Study, The University of Tokyo, Kashiwa, Chiba 277-8583, Japan}

\raggedbottom

\date{\today}

\begin{abstract}

Experiments searching for the electric dipole moment (EDM) of the electron $d_e$ utilise atomic/molecular states with one or more uncompensated electron 
spins, and these paramagnetic systems have recently achieved remarkable sensitivity to $d_e$.
If the source of $CP$ violation resides entirely in the hadronic sector, the two-photon exchange processes between electrons
and the nucleus induce $CP$-odd semileptonic interactions, parametrised by the Wilson coefficient $C_{SP}$, and provide the dominant source of EDMs in paramagnetic systems instead of $d_e$. 
We evaluate the $C_{SP}$ coefficients induced by the leading hadronic sources of $CP$ violation, namely nucleon EDMs and $CP$-odd pion-nucleon couplings, by calculating the nucleon-number-enhanced $CP$-odd nuclear scalar polarisability, employing chiral perturbation theory at the nucleon level and the Fermi-gas model for the nucleus. 
This allows us to translate the ACME EDM limits from paramagnetic ThO into novel independent constraints on the 
QCD theta term $|\bar \theta| < 3 \times 10^{-8}$, 
proton EDM $|d_p| < 2 \times 10^{-23}\,e\,{\rm cm}$, 
isoscalar $CP$-odd pion-nucleon coupling $|\bar g^{(1)}_{\pi NN}| < 4 \times 10^{-10}$, 
and colour EDMs of quarks $|\tilde d_u - \tilde d_d| < 2 \times 10^{-24}\,{\rm cm}$. 
We note that further experimental progress with EDM experiments in paramagnetic systems may allow them to rival the sensitivity of EDM experiments with neutrons and diamagnetic atoms to these quantities.  
\end{abstract}

\maketitle

\section{Introduction}
The origin of the matter-antimatter asymmetry in the universe continues to present us with one of the primary empirical motivations for new physics. The observed predominance of matter over anti-matter, quantified by the baryon-to-photon ratio $\et_B \approx 6\times 10^{-10}$, is now precisely determined through cosmological observations. A dynamical explanation for such an asymmetry, under the minimal assumption of the hot Big Bang model, requires sources of $CP$ violation as first laid out by Sakharov more than half a century ago \cite{Sakharov:1967dj}. The known sources of $CP$ violation in the Standard Model (SM) of particle physics -- the CKM phase and QCD vacuum angle (the latter consistent with zero to high accuracy) --  are insufficient for this task, providing empirical motivation for new sources of $CP$ violation that can explain the asymmetry. 

Experimental searches for new sources of $CP$- or $P,T$-violation have a long history, and electric dipole moments (EDMs) of nucleons, atoms and molecules are some of the primary observables (see, e.g., \cite{Khriplovich:1997ga,Ginges:2003qt,Pospelov:2005pr,Engel:2013lsa}). 
Fundamental sources of $T$ violation can feed into these observables in a variety of ways, depending on the specific features of the nuclear or atomic system. 
Searches for a neutron EDM pre-date even the discovery of $P$ violation in the weak interactions, but the use of atomic and now molecular systems to probe $CP$ violation through EDM-like observables has a more recent history, having been motivated by the possibility of various enhancement mechanisms \cite{Sandars,Flambaum1976_eEDM,Haxton:1983dq,Flambaum:1984fb}. 
Atomic/molecular EDM experiments are usually classified by whether the relevant state is either paramagnetic or diamagnetic, with the former principally aimed at probing leptonic sources of $CP$ violation via the EDMs of unpaired electrons, and the latter mainly probing hadronic sources of $CP$ violation via nuclear moments \cite{footnoteA}. 

The prevailing classification of EDM experiments has been valuable in assessing their complementarity in probing the full range of potential $CP$-violating sources. However, in recent years, there has been quite dramatic progress particularly in the use of paramagnetic molecular states to search for EDMs \cite{Hudson:2011zz,Baron:2013eja,Cairncross:2017fip,Andreev:2018ayy}. Indeed, the effective sensitivity to the electron EDM $d_e$ via atomic and now molecular EDM experiments has improved by a factor of more than $100$ over the past decade. This raises the question of whether EDM experiments in paramagnetic systems may soon provide significant sensitivity to hadronic sources of $CP$ violation, complementary to that from experiments targeting EDMs in diamagnetic systems and the nuclear Schiff moment \cite{Graner:2016ses,Xe-EDM_2019-Heil,Xe-EDM_2019-Fierlinger}, and also the neutron EDM \cite{Afach:2015sja}. 

In the present paper, we address this question quantitatively, by focusing on the sensitivity of paramagnetic EDM experiments, such as ACME \cite{Andreev:2018ayy}, to $CP$-odd semileptonic operators of the form, 
\be
\label{generic_contact_L}
 {\cal L} = C_{SP}^s \frac{G_F}{\sqrt{2}} \bar{e} i\gamma_5 e (\bar{p}p + \bar{n} n )+ C_{SP}^t \frac{G_F}{\sqrt{2}} \bar{e} i \gamma_5 e (\bar{p}p-\bar{n} n ) \, ,
\ee
where $e$, $n$ and $p$ refer to the electron, neutron and proton fields, respectively, and $C_{SP}^{s,t}$ are the couplings for the singlet and triplet operators, respectively. 
The subscript $SP$ denotes the nucleon-scalar and electron-pseudoscalar two-fermion bilinears. 
The semileptonic operators $C_{SP}$ in (\ref{generic_contact_L}) arise in the absence of any nuclear spin and are coherently enhanced by the number of nucleons in the nucleus, singling them out as the primary contributors to paramagnetic EDMs beyond the electron EDM, $-\frac{i}{2} d_e \bar{e} F_{\mu\nu} \sigma^{\mu\nu} \gamma_5 e$. 
Hadronic contributions to $d_e$, e.g.~from the QCD $\theta$ term, have been considered previously \cite{Choi:1990cn,Ghosh:2017uqq}, but the semileptonic operators above provide the leading sensitivity in atomic and molecular experiments. 
In particular, the leading source of paramagnetic EDMs due to the CKM phase is the $C_{SP}$ operator \cite{Pospelov:2013sca},
mediated by two-photon exchange. 
Beyond the Standard Model and extensions involving extra elementary-particle generations, new sources of $CP$ violation that manifest themselves in paramagnetic systems predominantly via the semileptonic operator $C_{SP}$, rather than $d_e$, may arise in supersymmetric models and multi-Higgs doublet models (for a general overview of these types of models, see e.g.~\cite{Pospelov:2005pr}).

In paramagnetic EDM experiments, the induced shift of atomic/molecular energy levels under an applied external electric field ${\cal E}_{\rm ext}$ can be written in the form 
\be
\label{DE}
\Delta E = - d_e {\cal E}_{\rm eff} - W_c \left[C^s_{SP} + \left(\frac{Z-N}{A}\right) C_{SP}^t \right] + \cdots  \, , 
\ee
where the factors ${\cal E}_{\rm eff}$ and $W_c$ are quantities that depend on the small ${\cal E}_{\rm ext}$, and $Z$, $N$ and $A$ denote the proton, neutron and total nucleon numbers of the nucleus, respectively. They are enhanced by a relativistic violation of the Schiff theorem and (for molecular systems) the polarisability \cite{Sandars}, and are now known to good precision for a variety of molecular species, see e.g.~\cite{PhysRevA.45.R4210,PhysRevA.80.062509,Mosyagin_1998,PhysRevA.84.052108,PhysRevA.78.010502,skripnikov2013communication,Jung:2013mg}. 
The existing null result from the ACME experiment \cite{Andreev:2018ayy}, using ThO, leads to the following 90\% confidence-level constraint on the effective $C_{SP}$ coupling averaged over the $p-n$ composition of the Th nucleus: 
\be
\label{CSs}
 |C_{SP}^s - 0.22 C_{SP}^t | = |0.39C_{SP}^p + 0.61 C_{SP}^n | < 7.3 \times 10^{-10} \, .   
\ee
Quite generically, for hadronic sources of $CP$ violation, the $d_e$ contribution to atomic/molecular EDMs is subdominant to $C_{SP}$.

\begin{figure}[t]
\centering
\includegraphics[width=3.95cm]{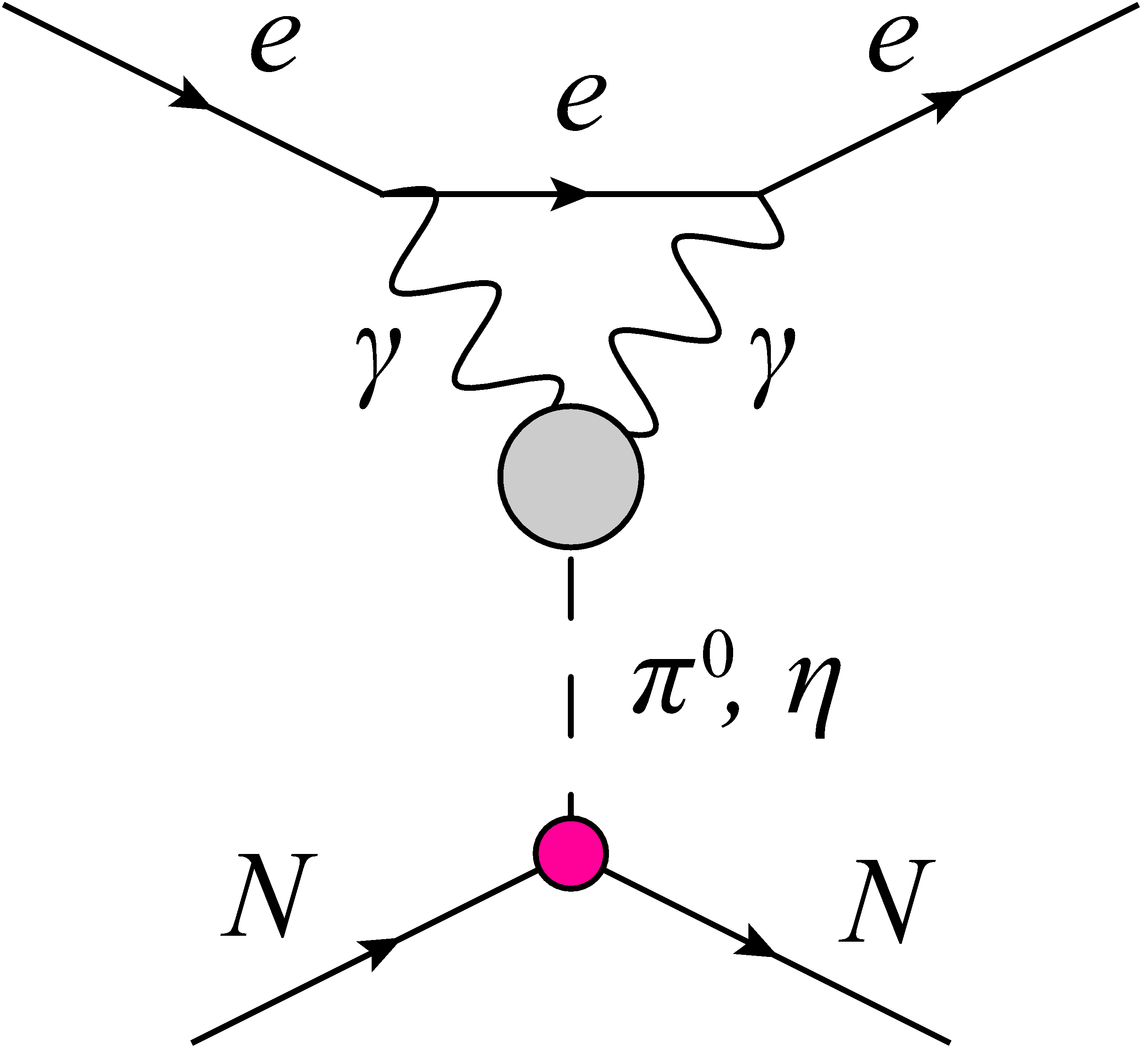}
\caption{(Color online)
$CP$-violating leading order (LO) semileptonic processes involving the exchange of a $\pi^0$ or $\eta$ meson. 
The grey vertex denotes the anomalous coupling (at the one-loop level) of the $\pi^0/\eta$ meson to the electromagnetic field, while the magenta vertex denotes the $CP$-violating coupling with the nucleon. 
}
\label{Fig:Semileptonic_theta_A}
\end{figure}

\begin{figure}[t]
\centering
\includegraphics[width=3.95cm]{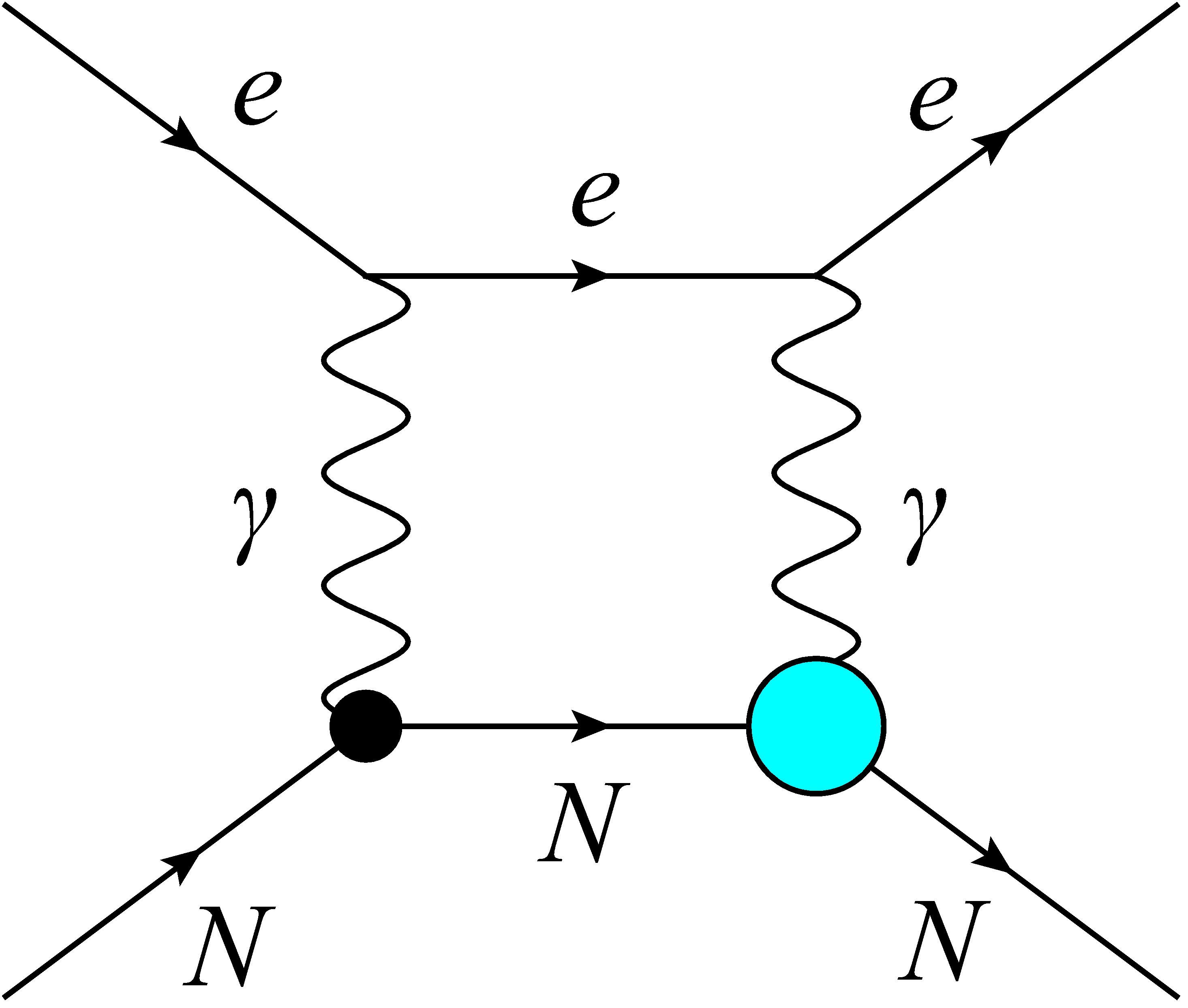}
\hspace{0.5cm}
\includegraphics[width=3.95cm]{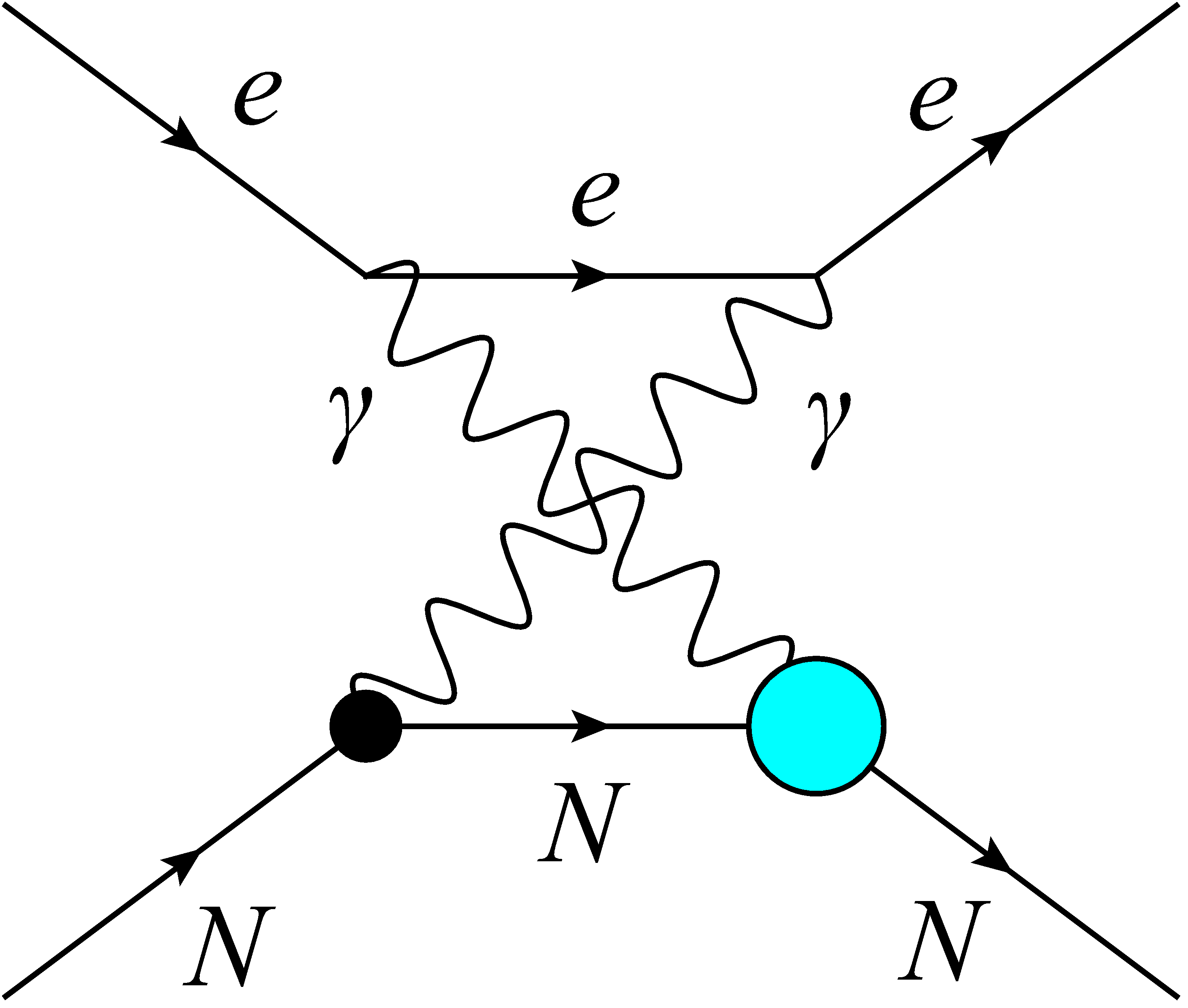}
\caption{(Color online)
$CP$-violating $\mu-d$ semileptonic processes with internal nuclear excitations. 
The black vertex again denotes the interaction of the electromagnetic field with the nucleon magnetic dipole moment $\mu$, while the cyan vertex denotes the interaction with the nucleon electric dipole moment $d$. 
The analogous processes with the black and cyan vertices interchanged are implicit. 
}
\label{Fig:Semileptonic_theta_BC}
\end{figure}

The semileptonic operators in (\ref{generic_contact_L}) can in turn be induced by the leading sources of $CP$ violation at the hadronic level, 
\begin{align}
  {\cal L}_{\rm hadronic}& = -\frac{i}{2} d_n \bar{n} F_{\mu\nu} \si^{\mu\nu} \gamma_5 n - \frac{i}{2} d_p\bar{p} F_{\mu\nu} \si^{\mu\nu} \gamma_5p \nonumber
  \\
   &\;\;\;\;+\bar{g}^{(0)}_{\pi NN} \bar{N} \ta^a N \pi^a + \bar{g}^{(1)}_{\pi NN} \bar{N} N \pi^0 +... \, ,  \label{Lhad}
\end{align}
where $N=(p,n)^T$ is the nucleon doublet, $d_{n,p}$ refers to nucleon EDMs, and $\bar{g}^{(0,1)}_{\pi NN}$ are the isovector and isoscalar $CP$-odd pion-nucleon couplings, respectively. 
This formula can also be generalised to include $CP$-odd interactions with the octet $\eta$ meson, $\eta \bar N(\bar g_{\eta NN}^{(0)} + \bar g_{\eta NN}^{(1)} \tau^3)N$. 
Thus we aim to determine
\be
 C_{SP} = C_{SP}(d_n,d_p,\bar{g}^{(0)}_{\pi/\et NN}, \bar{g}^{(1)}_{\pi/\et NN}, \ldots) \, ,
 \label{Cs}
\ee
that can be induced in particular by two-photon exchange processes (see Figs.~\ref{Fig:Semileptonic_theta_A}, \ref{Fig:Semileptonic_theta_A2} and \ref{Fig:Semileptonic_theta_BC}). 
The hadronic-scale interactions in (\ref{Lhad}) are in turn induced by more fundamental sources, such as $\theta_{\rm QCD}$, quark EDMs and chromo EDMs \cite{Pospelov:2005pr}. 
In what follows, we will examine the leading dependencies in (\ref{Cs}), and explore the induced sensitivity to fundamental $CP$-violating hadronic sources.

\section{Semileptonic operators induced by CP-odd nucleon polarisabilities}
When the underlying sources of $CP$ violation are hadronic and the nuclei of interest are spinless, the semileptonic couplings $C_{SP}$ in (\ref{generic_contact_L}) can be generated by two-photon exchange processes via $CP$-odd nucleon polarisabilities, 
\begin{align}
 {\cal L} &= -\frac{1}{4}\bar N (
 \beta_s + \tau^3 \beta_t)NF_{\mu\nu}\widetilde{F}^{\mu\nu} \label{betas} \\
 &= ( \beta_p \bar pp + \beta_n \bar nn ){\v{E} \cdot \v{B}} \, .  \label{betas-extra} 
\end{align}
Application of an external electric field $\v{E}$ leads to an {\it induced} magnetic dipole moment $\beta \v{E}$, and the sign in (\ref{betas},\ref{betas-extra}) is chosen to coincide with the $CP$-even polarisability convention, ${\cal L} = \alpha_\mathrm{pol}\v{E}^2 /2$.

\begin{figure}[t]
\centering
\includegraphics[width=4.05cm]{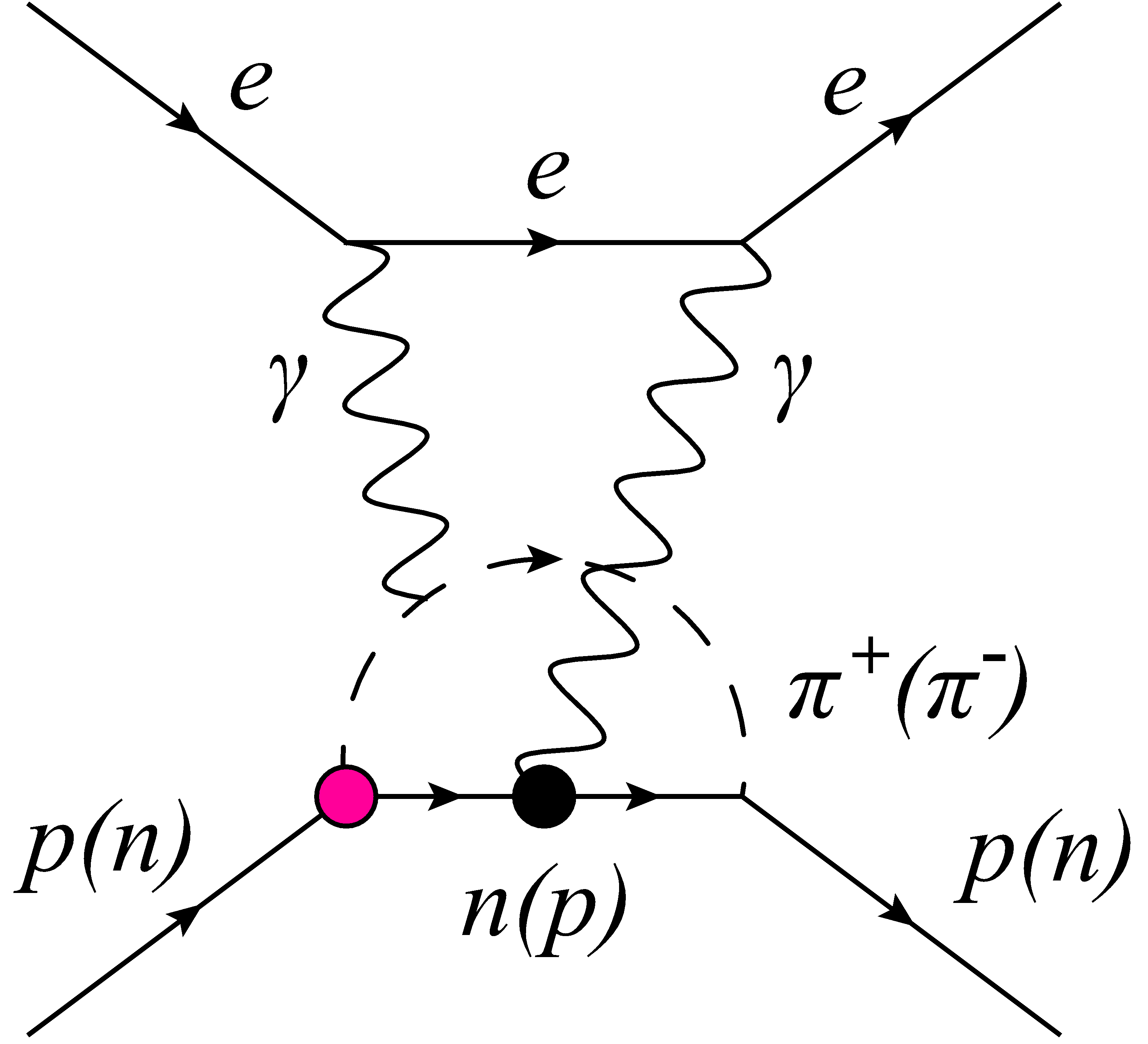}
\hspace{0.3cm}
\includegraphics[width=4.05cm]{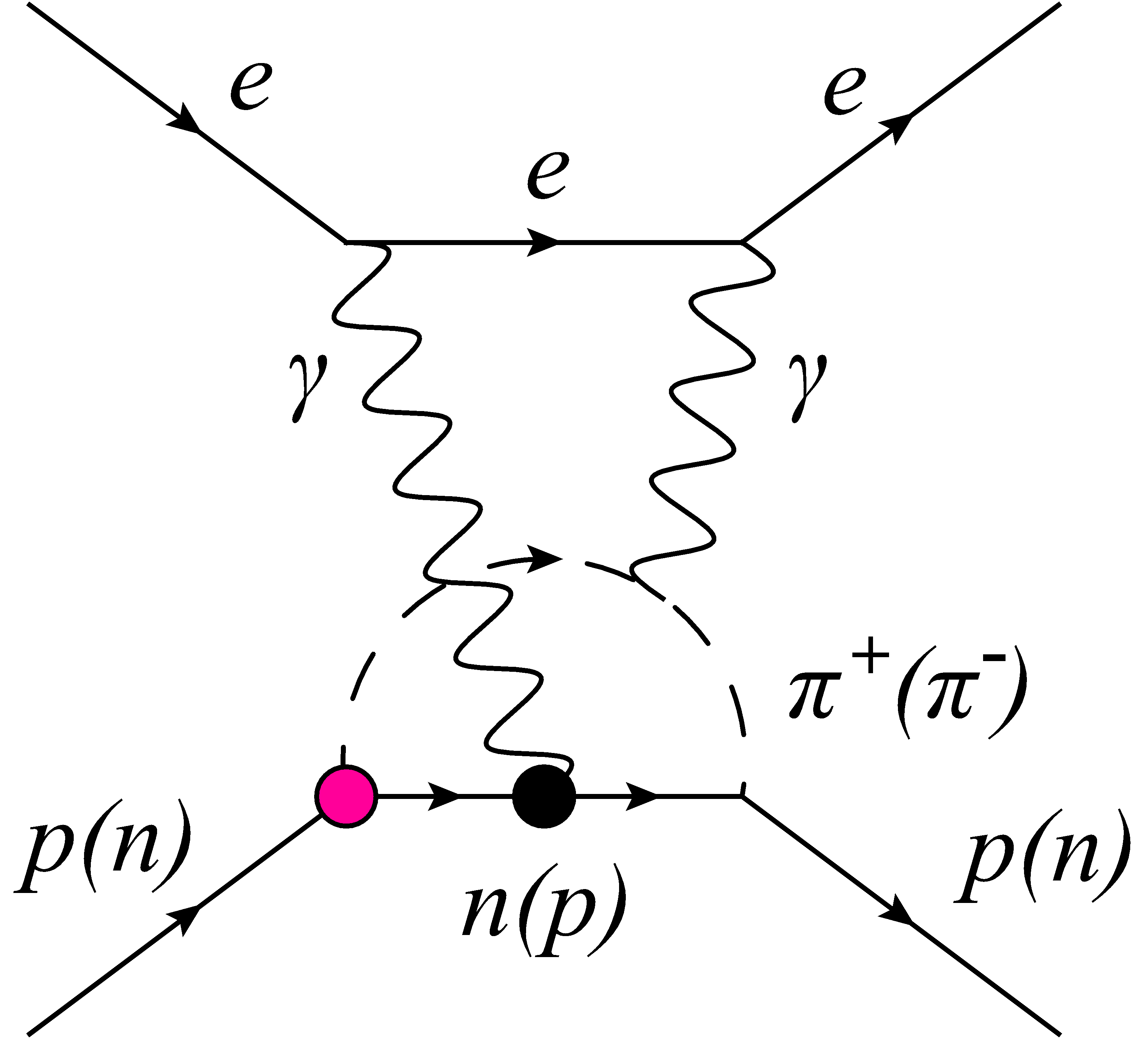}
\caption{(Color online)
$CP$-violating next-to-leading order (NLO) semileptonic processes involving a charged-pion loop. 
The magenta vertex again denotes the $CP$-violating coupling of the pion with the nucleon, while the black vertex denotes the coupling of the electromagnetic field to the nucleon magnetic dipole moment. 
The analogous processes with the magenta vertex interchanged with the other pion-nucleon vertex are implicit. 
}
\label{Fig:Semileptonic_theta_A2}
\end{figure}

A complete calculation of the $CP$-odd nuclear scalar polarisability is a complicated task, but at the nucleon level it can be performed using chiral perturbation 
theory. The leading order (LO) terms arise at ${\cal O}(m_\pi^{-2})$ in the pion mass $m_\pi$, as shown in Fig.~\ref{Fig:Semileptonic_theta_A}, and are given by 
\be
 \beta_{p(n)}^{LO}{=-}\frac{\al}{\pi F_\pi m_\pi^2} \left[\bar{g}^{(1)}_{\pi NN}{+(-)}\bar{g}^{(0)}_{\pi NN}{+}\frac{{\bar g}^{(0)}_{\eta NN}}{\sqrt{3}} \frac{m_\pi^2F_\pi}{m_\eta^2F_\eta} \right]  \, , 
\ee
where $F_\pi \approx 92$\,MeV is the pion decay constant, and $F_\eta$ is the octet $\eta$-meson decay constant, which we take to be $F_\eta \approx F_\pi$. 
The appearance of the factor $\alpha/\pi$ in this formula is due to the one-loop nature of the $\pi^0\gamma\gamma$ vertex.  We have neglected small isospin-breaking effects, $\eta-\eta'$  and $\pi^0-\eta$ mixings, as well as $\bar g_{\eta NN}^{(1)}$, as only the singlet contribution of $\eta$ proves to be important in the concrete examples below.
We next address the first formally sub-leading correction, which emerges from a charged-pion loop that interacts with $\v{E}$, while the magnetic moment of the nucleon interacts with $\v{B}$ (see Fig.~\ref{Fig:Semileptonic_theta_A2}). 
The next-to-leading order (NLO) result arises at ${\cal O}(m_\pi^{-1})$, and is given by
\be
 \beta^{NLO}_{k} = \frac{\alpha g_A\bar g^{(0)}_{\pi NN}}{4 F_\pi m_Nm_\pi }  
 \left\{ \begin{array}{cc} -\mu_n/\mu_N &{\rm for}\;k=p \, , \\  \mu_p/\mu_N &{\rm for}\;k=n \, ,\end{array}\right.
\ee
where $g_A \approx 1.3$ is the axial triplet coupling, $m_N$ is the nucleon mass, $\mu_{n,p}$ are the nucleon magnetic dipole moments, and $\mu_N$ is the nuclear magneton. 
We observe that this answer is numerically rather larger than would have naively been expected, in part as a result of the large values of $\mu_{n,p}$. 
Also, the $CP$-odd polarisabilities of neutrons and protons have the same sign, as $\mu_n$ is negative while $\mu_p$ is positive, and so add constructively.

To compute the contributions to $C_{SP}$, we next perform the integral over the diphoton loop, which is soft compared to the hadronic scales that were integrated out above, and average the result over the nucleon content in a nucleus. 
We find, to logarithmic accuracy, a known result for the semileptonic operator in the contact approximation: 
\begin{equation}
\label{log}
\frac{G_F}{\sqrt{2}} C_{SP}^{(\beta)} = -\left(\frac{Z}{A}\beta_p +\frac{N}{A}\beta_n\right) \frac{3\alpha m_e}{2\pi} \ln\left(\frac{\mathcal{M}}{m_e}\right)  \, .
\end{equation}
In the limit of a pointlike and structureless nucleus, the renormalisation scale $\mathcal{M}$ is {\it different} for the LO and NLO contributions:~for the LO terms, it is set by the $\pi$/$\eta$ form factor (i.e., a hadronic scale related to the $\rho$ meson mass $m_\rho$), while for the NLO process, $\mathcal{M} \approx m_\pi$ due to the presence of the pion propagators in the charged-pion loop. 
The nuclear size, which sets the value of the atomic $s-p$ mixing matrix element induced by $C_{SP}$ \cite{Bouchiat1975EDM,Khriplovich:1991mba}, does not play any role in regularising the integral, which extends down to $\sim m_e$ (corresponding to an interaction on the length scale $\sim m_e^{-1}$). 
The modification of the forms of relativistic atomic wavefunctions on the super-nuclear length scales $(8Z\alpha m_e)^{-1} \lesssim r \lesssim m_e^{-1}$ in sufficiently heavy atoms (see, e.g., \cite{Khriplovich:1991mba}) gives rise to non-logarithmic corrections to atomic $s-p$ mixing matrix elements. 
We also note that going beyond the logarithmic approximation in the NLO case would prevent the factorisation of the photon and pion loops, and would necessitate a full two-loop calculation. 

Thus far, we have neglected the fact that the internal nuclear dynamics may affect the values of the $\beta$ coefficients, and also lead to additional 
contributions to the $C_{SP}$ coefficients. 
For example, the pion loop calculation in the NLO process above assumed that the intermediate nucleon propagator is ``free'', 
while in reality it will be modified by nuclear in-medium effects. 
Moreover, EDMs of individual nucleons will lead to semileptonic operators 
that do not reduce to the simple $\v{E} \cdot \v{B}$ nuclear polarisability form --- we now address these types of processes.


\section{Semileptonic operators induced by nucleon EDMs}
Let us consider the semileptonic processes in Fig.~\ref{Fig:Semileptonic_theta_BC} that correspond to the exchange of two photons between atomic electrons and nucleons, with internal nuclear excitations. 
In this case, we assume that the nucleons possess both magnetic ($\mu$) and electric ($d$) dipole moments, as defined in (\ref{Lhad}) for the latter. 
We consider the simplest non-interacting Fermi-gas model of the nucleus, and single nucleon excitations. Integrating over the temporal component of the loop momentum leads to the following result per nucleon,
\begin{equation}
\label{2-photon_exchange_ampl_A}
\left.\frac{G_F}{\sqrt{2} }
C_{SP}^{(\mu d)}\right|_{\rm per~nucleon}  \approx \frac{4m_e \alpha \mu d}{3 \pi^2 } \int \frac{d^3 \v{k}}{\left| \v{k} \right|^4} \, , 
\end{equation}
where $\v{k}$ is the spatial part of the loop momentum. The integral in this case is dominated by the residue near the nucleon pole, with the virtual electron and photons deeply off-shell. 
Note also that Schiff's screening theorem \cite{PhysRev.132.2194} does not apply to the semileptonic processes under consideration, since the interaction of one of the virtual photons with the nucleon (or nucleus) is magnetic in nature.

To generalise to the case of a nucleus with $A \gg 1$ nucleons, we have to average the product $\mu d$ over the $p-n$ content of the nucleus, 
and evaluate the integral over the spatial loop momenta. 
If the initial nucleon momentum is $\v{p}$, where in the Fermi-gas model $|\v{p}|\leq p_F$,
then the intermediate nucleon momentum typically lies above the Fermi surface, $|\v{p} + \v{k} | \geq p_F$. 
This provides the IR regularisation of the integral in (\ref{2-photon_exchange_ampl_A}) that can be readily computed in terms of $p=|\v{p}|$ and the Fermi momentum $p_F$,
\begin{equation}
\label{loop_integral_IR_divergent}
\int \frac{d^3 \v{k}}{\left| \v{k} \right|^4} = \frac{4\pi}{p_F}\left[\frac{1}{2(1-x^2)} + \frac{1}{4x} \ln\left(\frac{1+x}{1-x}\right)\right]  \, ,~x=p/p_F  \, .
\end{equation}
Averaging over $x$ with the normalised $3x^2dx$ distribution in the interval $0<x<1$ leaves a logarithmic divergence as $x\to 1$ from the first term in (\ref{loop_integral_IR_divergent}). 
We regularise this average by taking into account the finite number of nucleons in a nucleus, $x_\textrm{max} \approx 1-1/(3A)$: 
\begin{equation}
\label{Fermi-gas_integral}
\frac{1}{4\pi}\left< \int \frac{d^3 \v{k}}{\left| \v{k} \right|^4} \right> \approx \frac{3}{4p_F} \ln \left( A \right)  \,  ,
\end{equation}
where $p_F \approx 250~\textrm{MeV}$ is the typical Fermi momentum of a nucleus. 
The contribution (\ref{2-photon_exchange_ampl_A}), generalised to a heavy nucleus, then takes the form: 
\begin{align}
C_{SP}^{(\mu d)}  &\approx  \frac{8 \sqrt{2} m_e \alpha^2  \ln(A)}{G_F p_F m_N} \left( \frac{Z}{A} \frac{\mu_p}{\mu_N} \frac{d_p}{e}  + \frac{N}{A} \frac{\mu_n}{\mu_N} \frac{d_n}{e} \right)  \label{2-photon_exchange_ampl_B} \\
&\approx  3.4 \times 10^{-11} \times \left(\frac{d_p-d_n}{10^{-24}\,e{\rm \, cm}}\right)  \, ,  \label{2-photon_exchange_ampl_B-extra}
\end{align}
where we have averaged the product $\langle \mu d \rangle$ over all nucleons in the nucleus. 
(In the second line above, we have presented an approximate expression for a heavy nucleus with $A \sim 200$ and $Z/A \sim 0.4$, for which the orthogonal combination $d_p + d_n$ can be neglected.) 


The result (\ref{2-photon_exchange_ampl_B}) does not exhaust all possible nuclear contributions. 
In particular, $CP$-violating
 hadronic sources will also induce contributions to the \textit{CP}-odd nuclear scalar polarisability from the near-outer-shell nucleons that will depend on the 
 details of the discrete nuclear structure. 
 The calculation of such effects goes beyond the scope of this work. 
 We believe that the bulk contribution of nucleons, which is enhanced in the atomic/molecular EDM by the factor ${\cal O}(A)$ and so grows with the total number of nucleons in a regular manner, is adequately captured by our treatment above.

\section{Constraints on CP-violating parameters}
 We now turn the current experimental limit on $C_{SP}$ given in (\ref{CSs}) into constraints on the
 fundamental parameters characterising $CP$ violation in the hadronic sector. 
Using the $\mu - d$ interactions of protons inside the nucleus (\ref{2-photon_exchange_ampl_B}), 
and neglecting $d_n$ which is already constrained directly, we derive the novel bound
\be
 |d_p|_{\rm ThO} < 2 \times 10^{-23}  \,  e \, {\rm cm}  \, ,
\ee
which is only a factor of 100 less stringent than the limit derived from the constraint on the Hg EDM \cite{Graner:2016ses}.

Next we address constraints related to the LO $\pi^0$ exchange, when the neutron and proton contributions add constructively, 
which is the case with the $\bar g^{(1)}_{\pi NN}$ coupling. We immediately obtain the limit
\begin{equation}
\label{g1limit}
|\bar g^{(1)}_{\pi NN}|_{\rm ThO} < 4 \times 10^{-10}  \, ,
\end{equation}
that upon use of the QCD sum-rule estimate of $\bar{g}^{(1)}_{\pi NN}(\tilde{d}_q)$ \cite{Pospelov:2001ys} can be translated into a limit on the isovector combination of light-quark colour EDMs (CEDMs):
\begin{equation}
\label{CEDM}
|\tilde d_u-\tilde d_d|_{\rm ThO} < 2\times 10^{-24} \, {\rm cm}  \, .
\end{equation}
These limits are more stringent than those derived from Xe EDM experiments \cite{Xe-EDM_2019-Heil,Xe-EDM_2019-Fierlinger} and are comparable to those derived from neutron EDM experiments \cite{nEDM-isovector_2014}, but are nominally less stringent than the limits derived from Hg EDM experiments using constraints on the nuclear Schiff moment \cite{Graner:2016ses}. 
However, the most recent nuclear calculations of the dependence of the Hg Schiff moment on the $\bar{g}^{(1)}_{\pi NN}$ coupling indicate significant sensitivity to assumptions about the underlying nuclear structure \cite{Ban:2010ea} (indeed the sensitivity of this system to $\bar{g}^{(1)}_{\pi NN}$ can be formally null within such nuclear uncertainties), which also propagates to substantial uncertainty in the limits on CEDMs inferred from the Hg EDM bound. 
Our result in (\ref{g1limit}) provides a completely independent limit on $\bar{g}^{(1)}_{\pi NN}$, with less theoretical uncertainty, since the effect is dominated by a bulk property of the nucleus.

We are now ready to address perhaps the most interesting quantity:~the limit on the QCD vacuum angle provided by EDM experiments in paramagnetic systems. 
To this end, we can utilise the chiral-limit results for the nucleon EDMs, $\bar{g}^{(0)}_{\pi NN}$ and $\bar g^{(0)}_{\eta NN}$
induced by $\bar \theta$:
\begin{align}
d_{p(n)} (\bar \theta) &= -(+) \frac{g_A\bar g^{(0)}_{\pi NN} e}{4\pi^2F_\pi}\ln\left(\frac{\mathcal{M}}{m_\pi}\right)  \, , \\
\bar{g}^{(0)}_{\pi NN} (\bar \theta) &=- \frac{m_* \bar\theta}{F_{\pi}} \langle p | \bar uu - \bar dd| p \rangle  \, , \\
 \bar g_{\eta NN}^{(0)} (\bar \theta) &= -\frac{m_* \bar\theta}{\sqrt{3}F_{\eta}} \langle p | \bar uu + \bar dd -2 \bar ss  | p \rangle  \, , 
\end{align}
where $m_* = m_um_d/(m_u+m_d)$, and the strange quark contribution to $m_*$ has been neglected. 
The renormalisation scale of the chiral loop \cite{Crewther:1979pi} can be taken to be $\mathcal{M} \sim 4 \pi F_\pi$, and the sub-logarithmic corrections have been neglected. 
[For a more in-depth treatment, one can use QCD sum-rule or lattice estimates of $d_N (\bar \theta)$.] 
The nucleon matrix elements are known to some accuracy from hadron spectroscopy and lattice calculations. 
Using $(m_d-m_u)\langle p |\bar uu -\bar dd |p\rangle \approx 2.5$ MeV, $(m_d+m_u)\langle p |\bar uu +\bar dd |p\rangle /2 \approx 38$ MeV and $\langle p |\bar ss |p\rangle \approx 0.1 \langle p |\bar uu +\bar dd |p\rangle$ \cite{Borsanyi:2014jba,Durr:2015dna}, one finds $\bar g^{(0)}_{\pi NN} \approx - 0.017 \bar \theta $, 
in good agreement with e.g.~\cite{chiral-PT_2015,Dragos:2019oxn}, and $\bar g^{(0)}_{\eta NN} \approx 5 \bar g^{(0)}_{\pi  NN}$. 
With these values, we observe that the LO contributions of $\pi^0$ and $\eta$ exchange to $C_{SP}$ almost cancel for the $p-n$ composition of the Th nucleus, as well as other heavy nuclei (but not light nuclei). 
Given the considerable degree of uncertainty in the quark bilinear matrix elements, this cancellation can suppress the naive $\pi^0$ exchange contribution by an order of magnitude or more, rendering the LO result intrinsically very uncertain in the case of heavy nuclei. 
However, we can combine the NLO contribution together with the $\mu - d $ contribution to obtain the following prediction for a heavy nucleus with $A \sim 200$ and $Z/A \sim 0.4$ (which includes nuclei of experimental interest such as Th, Tl, Hg, Hf and Xe): 
\begin{equation}
C_{SP}(\bar\theta) \approx  \left[0.1_\textrm{LO} + 1.0_\textrm{NLO} + 1.7_{(\mu d)} \right] \times 10^{-2} \bar\theta  \approx 0.03\,\bar\theta  \,  , 
\end{equation}
where the numbers in parentheses show the LO, NLO and $\mu-d$ contributions to $C_{SP}$, respectively. 
Each number here can vary by as much as $50\%$ (or more in the case of the LO contribution) upon varying $\mathcal{M}$ and other parameters. 
(We also note that the IR scale in the NLO contribution, $m_\pi$, 
can be renormalised somewhat inside the nucleus due, e.g., to Pauli blocking, and a shift in the in-medium value for $m_\pi$.) 
With these caveats, the above result translates to the following limit  on the QCD vacuum angle, 
\begin{equation}
|\bar \theta|_{\rm ThO} \lesssim 3\times 10^{-8}  \,  .
\end{equation}
This is only a factor of about 100 less stringent than the limit extracted from neutron EDM experiments.

\section{Discussion}
In this paper, we have shown that paramagnetic EDM experiments, by virtue of their dramatic recent gains, are now exhibiting levels of sensitivity to hadronic sources of $CP$ violation that are becoming competitive with experiments focusing directly on the nuclear Schiff moment and the neutron EDM. When the source of $CP$ violation is localised in the hadron sector, it is well known that the top-quark/Higgs two-loop mechanism can give a large contribution to $d_e$ \cite{Barr:1990vd}. On the other hand, as our paper demonstrates, 
when the main mediation mechanism is via light quarks, as is the case with the theta term and light-quark (C)EDMs, the main pathway for communicating $CP$ violation to the EDMs of paramagnetic systems is via the $C_{SP}$ operator in (\ref{generic_contact_L}), while $d_e$ can be neglected. 
This sensitivity arises through the two-photon generation of $C_{SP}$ that is coherently enhanced by the number of nucleons. We have considered two distinct two-photon exchange mechanisms for generating such $CP$-violating semileptonic operators:~(i) the exchange of $\pi^0$ and $\eta$ mesons between atomic electrons and nucleons, as well as charged-pion loops generating $CP$-odd nucleon polarisabilities, and (ii) $CP$-odd nuclear excitations due to nucleon EDMs.

\begin{table}[t]
\centering
\caption{
Summary of bounds on $CP$-violating hadronic parameters from the paramagnetic ThO EDM experiment derived in the present work, as well as from EDM experiments with neutrons and diamagnetic atoms. 
}
\label{tab:table-summary}
\setlength\extrarowheight{3pt}
\begin{tabular}{ |c|c|c|c|c| }%
\hline
System & $|d_p|$ ($e \cdot \textrm{cm}$) & $|\bar g^{(1)}_{\pi NN}|$ & $|\tilde d_u-\tilde d_d|$ (cm) & $|\bar{\theta}|$  \\ \hline 
\textbf{ThO} & $\mathbf{2 \times 10^{-23}}$ & $\mathbf{4 \times 10^{-10}}$ & $\mathbf{2 \times 10^{-24}}$ & $\mathbf{3 \times 10^{-8}}$  \\ \hline 
n & --- & $1.1 \times 10^{-10}$ & $5 \times 10^{-25}$ & $2.0 \times 10^{-10}$  \\ \hline 
Hg & $2.0 \times 10^{-25}$ & $1 \times 10^{-12}~^{\mathrm{\textbf{a}}}$ & $5 \times 10^{-27}~^{\mathrm{\textbf{a}}}$ & $1.5 \times 10^{-10}$  \\ \hline 
Xe & $3.2 \times 10^{-22}$ & $6.7 \times 10^{-8}$ & $3 \times 10^{-22}$ & $3.2 \times 10^{-6}$  \\ \hline 
\end{tabular}
\footnotetext{These limits can formally be null within nuclear uncertainties.} 
\end{table}

In Table~\ref{tab:table-summary}, we summarise our newly derived bounds from the paramagnetic ThO EDM experiment on the various $CP$-violating hadronic parameters and compare with bounds from EDM experiments with neutrons and diamagnetic atoms. 
The most precise result in our analysis is the constraint on the isoscalar $\bar g^{(1)}_{\pi NN}$ coupling, Eq.~(\ref{g1limit}), where the effect comes from $\pi^0$ exchange between unpaired electrons and the nucleus. 
This result is devoid of any substantial nuclear uncertainties, since the effect is dominated by a bulk property of the nucleus. 
When converted to a limit on light-quark CEDMs, the uncertainty is significant \cite{Pospelov:2001ys}, but future progress in lattice QCD calculations may reduce this substantially. 
The limits on other parameters, including $\bar \theta$, are sensitive to the assumptions about nuclear structure. 
We chose the simplest possible Fermi-gas model of the nucleus, exploiting the coherent nature of the effect, as $C_{SP}$ is contributed to by all nucleons inside a nucleus. 
We observe that for the $\mu-d$ contribution, there is a logarithmic enhancement, and the result (\ref{Fermi-gas_integral}) is also somewhat enhanced for the nucleon states close to the Fermi surface, which in turn are expected to be more sensitive to the details of the discrete nuclear structure. 
This suggests that our estimate of $C_{SP}^{(\mu d)}$ is probably correct only to within a factor of $\sim 2$, and would benefit from a more in-depth nuclear treatment. 
There are also reasons to suspect that the tree-level exchange of pions within a nucleus may provide some additional enhancement \cite{Flambaum:1984fb}
compared to the loop-level contributions considered here for the NLO and $\mu-d$ processes.

While generally less stringent by about two orders of magnitude than the limits inferred from EDM experiments in neutrons and diamagnetic atoms, the bounds derived via the novel mechanisms considered in this paper may soon become remarkably competitive. 
Future experimental progress with molecular EDM experiments, or potentially solid-state technologies (see, e.g., Refs. \cite{Heidenreich:2005sq,PhysRevA.73.022107,PhysRevLett.109.193003,Vutha:2018tsz}), may shrink this gap quite rapidly. 
We note that the novel mechanisms considered in the present paper arise even in the absence of nuclear spin, in principle enabling the utilisation of any nuclear isotope. This is in contrast to EDM experiments in diamagnetic atoms that rely on the mechanism of the nuclear Schiff moment, which can only arise for nuclei with nonzero spin. 
The results derived in this paper may also be used for searches of time-dependent $CP$-odd hadronic parameters via EDM experiments in paramagnetic systems, similarly to the analysis performed on $d_n$ in \cite{Abel:2017rtm} following the theoretical work of \cite{Graham2011axion,Stadnik2014axion}. 


\section*{Acknowledgments}
M.P.~is grateful to Dr.~A.~Dutt-Mazumder, his late friend for collaboration over a decade ago on unpublished work considering nucleon polarisabilities and EDMs. The authors also thank Dr.~M.~Voloshin for helpful discussions. The work of M.P.~and A.R.~was supported in part by NSERC, Canada.  
Research at the Perimeter Institute is supported in part by the Government of Canada through
NSERC and by the Province of Ontario through MEDT.
Y.V.S.~was supported by a Humboldt Research Fellowship from the Alexander von Humboldt Foundation, by the World Premier International Research Center Initiative (WPI), MEXT, Japan, and by the JSPS KAKENHI Grant Number 20K14460. 

\bibliographystyle{apsrev4-1.bst}
\bibliography{ParamagneticTheta}

\begin{thebibliography}{37}%
\makeatletter
\providecommand \@ifxundefined [1]{%
 \@ifx{#1\undefined}
}%
\providecommand \@ifnum [1]{%
 \ifnum #1\expandafter \@firstoftwo
 \else \expandafter \@secondoftwo
 \fi
}%
\providecommand \@ifx [1]{%
 \ifx #1\expandafter \@firstoftwo
 \else \expandafter \@secondoftwo
 \fi
}%
\providecommand \natexlab [1]{#1}%
\providecommand \enquote  [1]{``#1''}%
\providecommand \bibnamefont  [1]{#1}%
\providecommand \bibfnamefont [1]{#1}%
\providecommand \citenamefont [1]{#1}%
\providecommand \href@noop [0]{\@secondoftwo}%
\providecommand \href [0]{\begingroup \@sanitize@url \@href}%
\providecommand \@href[1]{\@@startlink{#1}\@@href}%
\providecommand \@@href[1]{\endgroup#1\@@endlink}%
\providecommand \@sanitize@url [0]{\catcode `\\12\catcode `\$12\catcode
  `\&12\catcode `\#12\catcode `\^12\catcode `\_12\catcode `\%12\relax}%
\providecommand \@@startlink[1]{}%
\providecommand \@@endlink[0]{}%
\providecommand \url  [0]{\begingroup\@sanitize@url \@url }%
\providecommand \@url [1]{\endgroup\@href {#1}{\urlprefix }}%
\providecommand \urlprefix  [0]{URL }%
\providecommand \Eprint [0]{\href }%
\providecommand \doibase [0]{http://dx.doi.org/}%
\providecommand \selectlanguage [0]{\@gobble}%
\providecommand \bibinfo  [0]{\@secondoftwo}%
\providecommand \bibfield  [0]{\@secondoftwo}%
\providecommand \translation [1]{[#1]}%
\providecommand \BibitemOpen [0]{}%
\providecommand \bibitemStop [0]{}%
\providecommand \bibitemNoStop [0]{.\EOS\space}%
\providecommand \EOS [0]{\spacefactor3000\relax}%
\providecommand \BibitemShut  [1]{\csname bibitem#1\endcsname}%
\let\auto@bib@innerbib\@empty
\bibitem [{\citenamefont {Sakharov}(1967)}]{Sakharov:1967dj}%
  \BibitemOpen
  \bibfield  {author} {\bibinfo {author} {\bibfnamefont {A.~D.}\ \bibnamefont
  {Sakharov}},\ }\href {\doibase 10.1070/PU1991v034n05ABEH002497} {\bibfield
  {journal} {\bibinfo  {journal} {Pisma Zh. Eksp. Teor. Fiz.}\ }\textbf
  {\bibinfo {volume} {5}},\ \bibinfo {pages} {32} (\bibinfo {year} {1967})},\
  \bibinfo {note} {[Usp.~Fiz.~Nauk~\textbf{161}, 61 (1991)]}\BibitemShut {NoStop}%
\bibitem [{\citenamefont {Khriplovich}\ and\ \citenamefont
  {Lamoreaux}(1997)}]{Khriplovich:1997ga}%
  \BibitemOpen
  \bibfield  {author} {\bibinfo {author} {\bibfnamefont {I.~B.}\ \bibnamefont
  {Khriplovich}}\ and\ \bibinfo {author} {\bibfnamefont {S.~K.}\ \bibnamefont
  {Lamoreaux}},\ }\href@noop {} {\emph {\bibinfo {title} {{CP violation without
  strangeness:~Electric dipole moments of particles, atoms, and molecules}}}},\
  (\bibinfo {year} {Springer, Heidelberg, 1997})\BibitemShut {NoStop}%
\bibitem [{\citenamefont {Ginges}\ and\ \citenamefont
  {Flambaum}(2004)}]{Ginges:2003qt}%
  \BibitemOpen
  \bibfield  {author} {\bibinfo {author} {\bibfnamefont {J.~S.~M.}\
  \bibnamefont {Ginges}}\ and\ \bibinfo {author} {\bibfnamefont {V.~V.}\
  \bibnamefont {Flambaum}},\ }\href {\doibase 10.1016/j.physrep.2004.03.005}
  {\bibfield  {journal} {\bibinfo  {journal} {Phys. Rept.}\ }\textbf {\bibinfo
  {volume} {397}},\ \bibinfo {pages} {63} (\bibinfo {year} {2004})},\ \Eprint
  {http://arxiv.org/abs/physics/0309054} {arXiv:physics/0309054 [physics]}
  \BibitemShut {NoStop}%
\bibitem [{\citenamefont {Pospelov}\ and\ \citenamefont
  {Ritz}(2005)}]{Pospelov:2005pr}%
  \BibitemOpen
  \bibfield  {author} {\bibinfo {author} {\bibfnamefont {M.}~\bibnamefont
  {Pospelov}}\ and\ \bibinfo {author} {\bibfnamefont {A.}~\bibnamefont
  {Ritz}},\ }\href {\doibase 10.1016/j.aop.2005.04.002} {\bibfield  {journal}
  {\bibinfo  {journal} {Annals Phys.}\ }\textbf {\bibinfo {volume} {318}},\
  \bibinfo {pages} {119} (\bibinfo {year} {2005})},\ \Eprint
  {http://arxiv.org/abs/hep-ph/0504231} {arXiv:hep-ph/0504231 [hep-ph]}
  \BibitemShut {NoStop}%
\bibitem [{\citenamefont {Engel}\ \emph {et~al.}(2013)\citenamefont {Engel},
  \citenamefont {Ramsey-Musolf},\ and\ \citenamefont {van
  Kolck}}]{Engel:2013lsa}%
  \BibitemOpen
  \bibfield  {author} {\bibinfo {author} {\bibfnamefont {J.}~\bibnamefont
  {Engel}}, \bibinfo {author} {\bibfnamefont {M.~J.}\ \bibnamefont
  {Ramsey-Musolf}}, \ and\ \bibinfo {author} {\bibfnamefont {U.}~\bibnamefont
  {van Kolck}},\ }\href {\doibase 10.1016/j.ppnp.2013.03.003} {\bibfield
  {journal} {\bibinfo  {journal} {Prog. Part. Nucl. Phys.}\ }\textbf {\bibinfo
  {volume} {71}},\ \bibinfo {pages} {21} (\bibinfo {year} {2013})},\ \Eprint
  {http://arxiv.org/abs/1303.2371} {arXiv:1303.2371 [nucl-th]} \BibitemShut
  {NoStop}%
\bibitem [{\citenamefont {Sandars}(1965)}]{Sandars}%
  \BibitemOpen
  \bibfield  {author} {\bibinfo {author} {\bibfnamefont {P.~G.~H.}\
  \bibnamefont {Sandars}},\ }\href@noop {} {\bibfield  {journal} {\bibinfo
  {journal} {Phys. Lett.}\ }\textbf {\bibinfo {volume} {14}},\ \bibinfo {pages}
  {194} (\bibinfo {year} {1965})}\BibitemShut {NoStop}%
\bibitem{Flambaum1976_eEDM}
V.~V.~Flambaum, Yad.~Fiz.~\textbf{24}, 383 (1976), [Sov.~J.~Nucl.~Phys.~\textbf{24}, 199 (1976)].\ 
\bibitem [{\citenamefont {Haxton}\ and\ \citenamefont
  {Henley}(1983)}]{Haxton:1983dq}%
  \BibitemOpen
  \bibfield  {author} {\bibinfo {author} {\bibfnamefont {W.~C.}\ \bibnamefont
  {Haxton}}\ and\ \bibinfo {author} {\bibfnamefont {E.~M.}\ \bibnamefont
  {Henley}},\ }\href {\doibase 10.1103/PhysRevLett.51.1937} {\bibfield
  {journal} {\bibinfo  {journal} {Phys. Rev. Lett.}\ }\textbf {\bibinfo
  {volume} {51}},\ \bibinfo {pages} {1937} (\bibinfo {year}
  {1983})}\BibitemShut {NoStop}%
\bibitem{Flambaum:1984fb}
V.~V.~Flambaum, I.~B.~Khriplovich, and O.~P.~Sushkov, Zh.~Eksp.~Teor.~Fiz.~\textbf{87}, 1521 (1984), [\href{http://jetp.ac.ru/cgi-bin/dn/e_060_05_0873.pdf}{Sov.~Phys.~JETP \textbf{60}, 873 (1984)}].\ 
\bibitem{footnoteA} For hadronic sources of $CP$ violation that manifest themselves in atoms and molecules via the nuclear Schiff moment, diamagnetic systems are generally advantageous compared with paramagnetic systems, due to the much smaller magnetic moments of diamagnetic systems that make them less susceptible to systematic effects of magnetic fields. 
Leptonic sources of $CP$ violation, such as the electron EDM, manifest themselves via uncompensated electron spins, and so efficiently manifest themselves in paramagnetic systems which possess one or more unpaired electron spins. 
The lack of unpaired electron spins in diamagnetic atoms and molecules, however, means that the effects of leptonic sources of $CP$ violation are strongly suppressed in diamagnetic systems, since the generation of an EDM in a diamagnetic system in this case arises in a higher order of perturbation theory via the magnetic hyperfine interaction \cite{Ginges:2003qt}. 
\bibitem [{\citenamefont {Hudson}\ \emph {et~al.}(2011)\citenamefont {Hudson},
  \citenamefont {Kara}, \citenamefont {Smallman}, \citenamefont {Sauer},
  \citenamefont {Tarbutt},\ and\ \citenamefont {Hinds}}]{Hudson:2011zz}%
  \BibitemOpen
  \bibfield  {author} {\bibinfo {author} {\bibfnamefont {J.~J.}\ \bibnamefont
  {Hudson}}, \bibinfo {author} {\bibfnamefont {D.~M.}\ \bibnamefont {Kara}},
  \bibinfo {author} {\bibfnamefont {I.~J.}\ \bibnamefont {Smallman}}, \bibinfo
  {author} {\bibfnamefont {B.~E.}\ \bibnamefont {Sauer}}, \bibinfo {author}
  {\bibfnamefont {M.~R.}\ \bibnamefont {Tarbutt}}, \ and\ \bibinfo {author}
  {\bibfnamefont {E.~A.}\ \bibnamefont {Hinds}},\ }\href {\doibase
  10.1038/nature10104} {\bibfield  {journal} {\bibinfo  {journal} {Nature}\
  }\textbf {\bibinfo {volume} {473}},\ \bibinfo {pages} {493} (\bibinfo {year}
  {2011})}\BibitemShut {NoStop}%
\bibitem [{\citenamefont {Baron}\ \emph {et~al.}(2014)\citenamefont {Baron}
  \emph {et~al.}}]{Baron:2013eja}%
  \BibitemOpen
  \bibfield  {author} {\bibinfo {author} {\bibfnamefont {J.}~\bibnamefont
  {Baron}} \emph {et~al.} (\bibinfo {collaboration} {ACME}),\ }\href {\doibase
  10.1126/science.1248213} {\bibfield  {journal} {\bibinfo  {journal}
  {Science}\ }\textbf {\bibinfo {volume} {343}},\ \bibinfo {pages} {269}
  (\bibinfo {year} {2014})},\ \Eprint {http://arxiv.org/abs/1310.7534}
  {arXiv:1310.7534 [physics.atom-ph]} \BibitemShut {NoStop}%
\bibitem [{\citenamefont {Cairncross}\ \emph {et~al.}(2017)\citenamefont
  {Cairncross}, \citenamefont {Gresh}, \citenamefont {Grau}, \citenamefont
  {Cossel}, \citenamefont {Roussy}, \citenamefont {Ni}, \citenamefont {Zhou},
  \citenamefont {Ye},\ and\ \citenamefont {Cornell}}]{Cairncross:2017fip}%
  \BibitemOpen
  \bibfield  {author} {\bibinfo {author} {\bibfnamefont {W.~B.}\ \bibnamefont
  {Cairncross}}, \bibinfo {author} {\bibfnamefont {D.~N.}\ \bibnamefont
  {Gresh}}, \bibinfo {author} {\bibfnamefont {M.}~\bibnamefont {Grau}},
  \bibinfo {author} {\bibfnamefont {K.~C.}\ \bibnamefont {Cossel}}, \bibinfo
  {author} {\bibfnamefont {T.~S.}\ \bibnamefont {Roussy}}, \bibinfo {author}
  {\bibfnamefont {Y.}~\bibnamefont {Ni}}, \bibinfo {author} {\bibfnamefont
  {Y.}~\bibnamefont {Zhou}}, \bibinfo {author} {\bibfnamefont {J.}~\bibnamefont
  {Ye}}, \ and\ \bibinfo {author} {\bibfnamefont {E.~A.}\ \bibnamefont
  {Cornell}},\ }\href {\doibase 10.1103/PhysRevLett.119.153001} {\bibfield
  {journal} {\bibinfo  {journal} {Phys. Rev. Lett.}\ }\textbf {\bibinfo
  {volume} {119}},\ \bibinfo {pages} {153001} (\bibinfo {year} {2017})},\
  \Eprint {http://arxiv.org/abs/1704.07928} {arXiv:1704.07928
  [physics.atom-ph]} \BibitemShut {NoStop}%
\bibitem [{\citenamefont {Andreev}\ \emph {et~al.}(2018)\citenamefont {Andreev}
  \emph {et~al.}}]{Andreev:2018ayy}%
  \BibitemOpen
  \bibfield  {author} {\bibinfo {author} {\bibfnamefont {V.}~\bibnamefont
  {Andreev}} \emph {et~al.} (\bibinfo {collaboration} {ACME}),\ }\href
  {\doibase 10.1038/s41586-018-0599-8} {\bibfield  {journal} {\bibinfo
  {journal} {Nature}\ }\textbf {\bibinfo {volume} {562}},\ \bibinfo {pages}
  {355} (\bibinfo {year} {2018})}\BibitemShut {NoStop}%
\bibitem [{\citenamefont {Graner}\ \emph {et~al.}(2016)\citenamefont {Graner},
  \citenamefont {Chen}, \citenamefont {Lindahl},\ and\ \citenamefont
  {Heckel}}]{Graner:2016ses}%
  \BibitemOpen
  \bibfield  {author} {\bibinfo {author} {\bibfnamefont {B.}~\bibnamefont
  {Graner}}, \bibinfo {author} {\bibfnamefont {Y.}~\bibnamefont {Chen}},
  \bibinfo {author} {\bibfnamefont {E.~G.}\ \bibnamefont {Lindahl}}, \ and\
  \bibinfo {author} {\bibfnamefont {B.~R.}\ \bibnamefont {Heckel}},\ }\href
  {\doibase 10.1103/PhysRevLett.119.119901, 10.1103/PhysRevLett.116.161601}
  {\bibfield  {journal} {\bibinfo  {journal} {Phys. Rev. Lett.}\ }\textbf
  {\bibinfo {volume} {116}},\ \bibinfo {pages} {161601} (\bibinfo {year}
  {2016})},\ \bibinfo {note} {[Erratum:~Phys.~Rev.~Lett.~\textbf{119}, 119901 (2017)]},\ \Eprint {http://arxiv.org/abs/1601.04339}
  {arXiv:1601.04339 [physics.atom-ph]} \BibitemShut {NoStop}%
\bibitem{Xe-EDM_2019-Heil}
F.~Allmendinger, I.~Engin, W.~Heil, S.~Karpuk, H.-J.~Krause, B.~Niederlander, A.~Offenhausser, M.~Repetto, U.~Schmidt, and S.~Zimmer, \href{\doibase 10.1103/PhysRevA.100.022505}{Phys.~Rev.~A \textbf{100}, 022505 (2019)},\ 
\Eprint {https://arxiv.org/abs/1904.12295} {arXiv:1904.12295 [physics.atom-ph]}. 
\bibitem{Xe-EDM_2019-Fierlinger}
N.~Sachdeva  \emph {et~al.}, \href{\doibase 10.1103/PhysRevLett.123.143003}{Phys.~Rev.~Lett.~\textbf{123}, 143003 (2019)},\ 
\Eprint {https://arxiv.org/abs/1902.02864} {arXiv:1902.02864 [physics.atom-ph]}. 
\bibitem [{\citenamefont {Pendlebury}\ \emph {et~al.}(2015)\citenamefont
  {Pendlebury} \emph {et~al.}}]{Afach:2015sja}%
  \BibitemOpen
  \bibfield  {author} {\bibinfo {author} {\bibfnamefont {J.~M.}\ \bibnamefont
  {Pendlebury}} \emph {et~al.},\ }\href {\doibase 10.1103/PhysRevD.92.092003}
  {\bibfield  {journal} {\bibinfo  {journal} {Phys. Rev. D}\ }\textbf {\bibinfo
  {volume} {92}},\ \bibinfo {pages} {092003} (\bibinfo {year} {2015})},\
  \Eprint {http://arxiv.org/abs/1509.04411} {arXiv:1509.04411 [hep-ex]}
  \BibitemShut {NoStop}%
\bibitem [{\citenamefont {Choi}\ and\ \citenamefont
  {Hong}(1991)}]{Choi:1990cn}%
  \BibitemOpen
  \bibfield  {author} {\bibinfo {author} {\bibfnamefont {K.}~\bibnamefont
  {Choi}}\ and\ \bibinfo {author} {\bibfnamefont {J.-Y.}\ \bibnamefont
  {Hong}},\ }\href {\doibase 10.1016/0370-2693(91)90838-H} {\bibfield
  {journal} {\bibinfo  {journal} {Phys. Lett.}\ }\textbf {\bibinfo {volume}
  {B259}},\ \bibinfo {pages} {340} (\bibinfo {year} {1991})}\BibitemShut
  {NoStop}%
\bibitem [{\citenamefont {Ghosh}\ and\ \citenamefont
  {Sato}(2018)}]{Ghosh:2017uqq}%
  \BibitemOpen
  \bibfield  {author} {\bibinfo {author} {\bibfnamefont {D.}~\bibnamefont
  {Ghosh}}\ and\ \bibinfo {author} {\bibfnamefont {R.}~\bibnamefont {Sato}},\
  }\href {\doibase 10.1016/j.physletb.2017.12.052} {\bibfield  {journal}
  {\bibinfo  {journal} {Phys. Lett.}\ }\textbf {\bibinfo {volume} {B777}},\
  \bibinfo {pages} {335} (\bibinfo {year} {2018})},\ \Eprint
  {http://arxiv.org/abs/1709.05866} {arXiv:1709.05866 [hep-ph]} \BibitemShut
  {NoStop}%
\bibitem [{\citenamefont {Pospelov}\ and\ \citenamefont
  {Ritz}(2014)}]{Pospelov:2013sca}%
  \BibitemOpen
  \bibfield  {author} {\bibinfo {author} {\bibfnamefont {M.}~\bibnamefont
  {Pospelov}}\ and\ \bibinfo {author} {\bibfnamefont {A.}~\bibnamefont
  {Ritz}},\ }\href {\doibase 10.1103/PhysRevD.89.056006} {\bibfield  {journal}
  {\bibinfo  {journal} {Phys. Rev. D}\ }\textbf {\bibinfo {volume} {89}},\
  \bibinfo {pages} {056006} (\bibinfo {year} {2014})},\ \Eprint
  {http://arxiv.org/abs/1311.5537} {arXiv:1311.5537 [hep-ph]} \BibitemShut
  {NoStop}%
\bibitem [{\citenamefont {Liu}\ and\ \citenamefont
  {Kelly}(1992)}]{PhysRevA.45.R4210}%
  \BibitemOpen
  \bibfield  {author} {\bibinfo {author} {\bibfnamefont {Z.~W.}\ \bibnamefont
  {Liu}}\ and\ \bibinfo {author} {\bibfnamefont {H.~P.}\ \bibnamefont
  {Kelly}},\ }\href {\doibase 10.1103/PhysRevA.45.R4210} {\bibfield  {journal}
  {\bibinfo  {journal} {Phys. Rev. A}\ }\textbf {\bibinfo {volume} {45}},\
  \bibinfo {pages} {R4210} (\bibinfo {year} {1992})}\BibitemShut {NoStop}%
\bibitem [{\citenamefont {Mosyagin}\ \emph {et~al.}(1998)\citenamefont
  {Mosyagin}, \citenamefont {Kozlov},\ and\ \citenamefont
  {Titov}}]{Mosyagin_1998}%
  \BibitemOpen
  \bibfield  {author} {\bibinfo {author} {\bibfnamefont {N.~S.}\ \bibnamefont
  {Mosyagin}}, \bibinfo {author} {\bibfnamefont {M.~G.}\ \bibnamefont
  {Kozlov}}, \ and\ \bibinfo {author} {\bibfnamefont {A.~V.}\ \bibnamefont
  {Titov}},\ }\href {\doibase 10.1088/0953-4075/31/19/002} {\bibfield
  {journal} {\bibinfo  {journal} {J.~Phys.~B}\ }\textbf {\bibinfo {volume} {31}},\ \bibinfo {pages} {L763}
  (\bibinfo {year} {1998})}\BibitemShut {NoStop}%
\bibitem [{\citenamefont {Meyer}\ and\ \citenamefont
  {Bohn}(2008)}]{PhysRevA.78.010502}%
  \BibitemOpen
  \bibfield  {author} {\bibinfo {author} {\bibfnamefont {E.~R.}\ \bibnamefont
  {Meyer}}\ and\ \bibinfo {author} {\bibfnamefont {J.~L.}\ \bibnamefont
  {Bohn}},\ }\href {\doibase 10.1103/PhysRevA.78.010502} {\bibfield  {journal}
  {\bibinfo  {journal} {Phys. Rev. A}\ }\textbf {\bibinfo {volume} {78}},\
  \bibinfo {pages} {010502} (\bibinfo {year} {2008})},\ 
\Eprint {https://arxiv.org/abs/0805.0161} {arXiv:0805.0161 [physics.atom-ph]}
\BibitemShut {NoStop}%
\bibitem [{\citenamefont {Dzuba}\ and\ \citenamefont
  {Flambaum}(2009)}]{PhysRevA.80.062509}%
  \BibitemOpen
  \bibfield  {author} {\bibinfo {author} {\bibfnamefont {V.~A.}\ \bibnamefont
  {Dzuba}}\ and\ \bibinfo {author} {\bibfnamefont {V.~V.}\ \bibnamefont
  {Flambaum}},\ }\href {\doibase 10.1103/PhysRevA.80.062509} {\bibfield
  {journal} {\bibinfo  {journal} {Phys. Rev. A}\ }\textbf {\bibinfo {volume}
  {80}},\ \bibinfo {pages} {062509} (\bibinfo {year} {2009})},\ 
\Eprint {https://arxiv.org/abs/0909.0308} {arXiv:0909.0308 [physics.atom-ph]}
\BibitemShut
  {NoStop}%
\bibitem [{\citenamefont {Dzuba}\ \emph {et~al.}(2011)\citenamefont {Dzuba},
  \citenamefont {Flambaum},\ and\ \citenamefont
  {Harabati}}]{PhysRevA.84.052108}%
  \BibitemOpen
  \bibfield  {author} {\bibinfo {author} {\bibfnamefont {V.~A.}\ \bibnamefont
  {Dzuba}}, \bibinfo {author} {\bibfnamefont {V.~V.}\ \bibnamefont {Flambaum}},
  \ and\ \bibinfo {author} {\bibfnamefont {C.}~\bibnamefont {Harabati}},\
  }\href {\doibase 10.1103/PhysRevA.84.052108} {\bibfield  {journal} {\bibinfo
  {journal} {Phys. Rev. A}\ }\textbf {\bibinfo {volume} {84}},\ \bibinfo
  {pages} {052108} (\bibinfo {year} {2011})},\ 
\Eprint {https://arxiv.org/abs/1109.6082} {arXiv:1109.6082 [physics.atom-ph]}  \BibitemShut {NoStop}%
\bibitem{skripnikov2013communication}
L.~V.~Skripnikov, A.~N.~Petrov, and A.~V.~Titov, \href{\doibase 10.1063/1.4843955}{J.~Chem.~Phys.~\textbf{139}, 221103 (2013)},\ 
\Eprint {https://arxiv.org/abs/1308.0414} {arXiv:1308.0414 [physics.chem-ph]}. 
\bibitem [{\citenamefont {Jung}(2013)}]{Jung:2013mg}%
  \BibitemOpen
  \bibfield  {author} {\bibinfo {author} {\bibfnamefont {M.}~\bibnamefont
  {Jung}},\ }\href {\doibase 10.1007/JHEP05(2013)168} {\bibfield  {journal}
  {\bibinfo  {journal} {JHEP}\ }\textbf {\bibinfo {volume} {05}},\ \bibinfo
  {pages} {168} (\bibinfo {year} {2013})},\ \Eprint
  {http://arxiv.org/abs/1301.1681} {arXiv:1301.1681 [hep-ph]} \BibitemShut
  {NoStop}%
\bibitem{Bouchiat1975EDM}
C.~Bouchiat, \href{\doibase 10.1016/0370-2693(75)90077-5}{Phys.~Lett.~B \textbf{57}, 284 (1975)}.\ 
\bibitem [{\citenamefont {Khriplovich}(1991)}]{Khriplovich:1991mba}%
  \BibitemOpen
  \bibfield  {author} {\bibinfo {author} {\bibfnamefont {I.~B.}\ \bibnamefont
  {Khriplovich}},\ }\href@noop {} {\emph {\bibinfo {title} {{Parity
  Nonconservation in Atomic Phenomena}}}},\ (\bibinfo  {publisher} {Gordon and
  Breach, Philadelphia},\ \bibinfo {year} {1991})\BibitemShut {NoStop}%
\bibitem [{\citenamefont {Schiff}(1963)}]{PhysRev.132.2194}%
  \BibitemOpen
  \bibfield  {author} {\bibinfo {author} {\bibfnamefont {L.~I.}\ \bibnamefont
  {Schiff}},\ }\href {\doibase 10.1103/PhysRev.132.2194} {\bibfield  {journal}
  {\bibinfo  {journal} {Phys. Rev.}\ }\textbf {\bibinfo {volume} {132}},\
  \bibinfo {pages} {2194} (\bibinfo {year} {1963})}\BibitemShut {NoStop}%
\bibitem [{\citenamefont {Pospelov}(2002)}]{Pospelov:2001ys}%
  \BibitemOpen
  \bibfield  {author} {\bibinfo {author} {\bibfnamefont {M.}~\bibnamefont
  {Pospelov}},\ }\href {\doibase 10.1016/S0370-2693(02)01263-7} {\bibfield
  {journal} {\bibinfo  {journal} {Phys. Lett.}\ }\textbf {\bibinfo {volume}
  {B530}},\ \bibinfo {pages} {123} (\bibinfo {year} {2002})},\ \Eprint
  {http://arxiv.org/abs/hep-ph/0109044} {arXiv:hep-ph/0109044 [hep-ph]}
  \BibitemShut {NoStop}%
\bibitem{nEDM-isovector_2014}
C.-Y.~Seng, J.~de Vries, E.~Mereghetti, H.~H.~Patel, and M.~Ramsey-Musolf, \href{\doibase 10.1016/j.physletb.2014.07.014}{Phys.~Lett.~B \textbf{736}, 147 (2014)},\ 
\Eprint {https://arxiv.org/abs/1401.5366} {arXiv:1401.5366 [nucl-th]}. 
\bibitem [{\citenamefont {Ban}\ \emph {et~al.}(2010)\citenamefont {Ban},
  \citenamefont {Dobaczewski}, \citenamefont {Engel},\ and\ \citenamefont
  {Shukla}}]{Ban:2010ea}%
  \BibitemOpen
  \bibfield  {author} {\bibinfo {author} {\bibfnamefont {S.}~\bibnamefont
  {Ban}}, \bibinfo {author} {\bibfnamefont {J.}~\bibnamefont {Dobaczewski}},
  \bibinfo {author} {\bibfnamefont {J.}~\bibnamefont {Engel}}, \ and\ \bibinfo
  {author} {\bibfnamefont {A.}~\bibnamefont {Shukla}},\ }\href {\doibase
  10.1103/PhysRevC.82.015501} {\bibfield  {journal} {\bibinfo  {journal} {Phys.
  Rev. C}\ }\textbf {\bibinfo {volume} {82}},\ \bibinfo {pages} {015501}
  (\bibinfo {year} {2010})},\ \Eprint {http://arxiv.org/abs/1003.2598}
  {arXiv:1003.2598 [nucl-th]} \BibitemShut {NoStop}%
\bibitem [{\citenamefont {Crewther}\ \emph {et~al.}(1979)\citenamefont
  {Crewther}, \citenamefont {Di~Vecchia}, \citenamefont {Veneziano},\ and\
  \citenamefont {Witten}}]{Crewther:1979pi}%
  \BibitemOpen
  \bibfield  {author} {\bibinfo {author} {\bibfnamefont {R.~J.}\ \bibnamefont
  {Crewther}}, \bibinfo {author} {\bibfnamefont {P.}~\bibnamefont
  {Di~Vecchia}}, \bibinfo {author} {\bibfnamefont {G.}~\bibnamefont
  {Veneziano}}, \ and\ \bibinfo {author} {\bibfnamefont {E.}~\bibnamefont
  {Witten}},\ }\href {\doibase 10.1016/0370-2693(80)91025-4,
  10.1016/0370-2693(79)90128-X} {\bibfield  {journal} {\bibinfo  {journal}
  {Phys. Lett.}\ }\textbf {\bibinfo {volume} {88B}},\ \bibinfo {pages} {123}
  (\bibinfo {year} {1979})},\ \bibinfo {note} {[Erratum:~Phys.~Lett.~\textbf{91B}, 487 (1980)]}\BibitemShut {NoStop}%
\bibitem [{\citenamefont {Borsanyi}\ \emph {et~al.}(2015)\citenamefont
  {Borsanyi} \emph {et~al.}}]{Borsanyi:2014jba}%
  \BibitemOpen
  \bibfield  {author} {\bibinfo {author} {\bibfnamefont {S.}~\bibnamefont
  {Borsanyi}} \emph {et~al.},\ }\href {\doibase 10.1126/science.1257050}
  {\bibfield  {journal} {\bibinfo  {journal} {Science}\ }\textbf {\bibinfo
  {volume} {347}},\ \bibinfo {pages} {1452} (\bibinfo {year} {2015})},\ \Eprint
  {http://arxiv.org/abs/1406.4088} {arXiv:1406.4088 [hep-lat]} \BibitemShut
  {NoStop}%
\bibitem [{\citenamefont {Durr}\ \emph {et~al.}(2016)\citenamefont {Durr} \emph
  {et~al.}}]{Durr:2015dna}%
  \BibitemOpen
  \bibfield  {author} {\bibinfo {author} {\bibfnamefont {S.}~\bibnamefont
  {Durr}} \emph {et~al.},\ }\href {\doibase 10.1103/PhysRevLett.116.172001}
  {\bibfield  {journal} {\bibinfo  {journal} {Phys. Rev. Lett.}\ }\textbf
  {\bibinfo {volume} {116}},\ \bibinfo {pages} {172001} (\bibinfo {year}
  {2016})},\ \Eprint {http://arxiv.org/abs/1510.08013} {arXiv:1510.08013
  [hep-lat]} \BibitemShut {NoStop}%
\bibitem{chiral-PT_2015}
J.~de Vries, E.~Mereghetti, and A.~Walker-Loud, \href{\doibase 10.1103/PhysRevC.92.045201}{Phys.~Rev.~C~\textbf{92}, 045201 (2015)},\ 
\Eprint {https://arxiv.org/abs/1506.06247}  {arXiv:1506.06247 [nucl-th]}. 
\bibitem [{\citenamefont {Dragos}\ \emph {et~al.}(2019)\citenamefont {Dragos},
  \citenamefont {Luu}, \citenamefont {Shindler}, \citenamefont {de~Vries},\
  and\ \citenamefont {Yousif}}]{Dragos:2019oxn}%
  \BibitemOpen
  \bibfield  {author} {\bibinfo {author} {\bibfnamefont {J.}~\bibnamefont
  {Dragos}}, \bibinfo {author} {\bibfnamefont {T.}~\bibnamefont {Luu}},
  \bibinfo {author} {\bibfnamefont {A.}~\bibnamefont {Shindler}}, \bibinfo
  {author} {\bibfnamefont {J.}~\bibnamefont {de~Vries}}, \ and\ \bibinfo
  {author} {\bibfnamefont {A.}~\bibnamefont {Yousif}},\ }\href@noop {} {\
  (\bibinfo {year} {2019})},\ \Eprint {http://arxiv.org/abs/1902.03254}
  {arXiv:1902.03254 [hep-lat]} \BibitemShut {NoStop}%
\bibitem [{\citenamefont {Barr}\ and\ \citenamefont {Zee}(1990)}]{Barr:1990vd}%
  \BibitemOpen
  \bibfield  {author} {\bibinfo {author} {\bibfnamefont {S.~M.}\ \bibnamefont
  {Barr}}\ and\ \bibinfo {author} {\bibfnamefont {A.}~\bibnamefont {Zee}},\
  }\href {\doibase 10.1103/PhysRevLett.65.2920, 10.1103/PhysRevLett.65.21}
  {\bibfield  {journal} {\bibinfo  {journal} {Phys. Rev. Lett.}\ }\textbf
  {\bibinfo {volume} {65}},\ \bibinfo {pages} {21} (\bibinfo {year} {1990})},\
  \bibinfo {note} {[Erratum:~Phys.~Rev.~Lett.~\textbf{65}, 2920 (1990)]}\BibitemShut
  {NoStop}%
\bibitem [{\citenamefont {Heidenreich}\ \emph {et~al.}(2005)\citenamefont
  {Heidenreich} \emph {et~al.}}]{Heidenreich:2005sq}%
  \BibitemOpen
  \bibfield  {author} {\bibinfo {author} {\bibfnamefont {B.~J.}\ \bibnamefont
  {Heidenreich}} \emph {et~al.},\ }\href {\doibase
  10.1103/PhysRevLett.95.253004} {\bibfield  {journal} {\bibinfo  {journal}
  {Phys. Rev. Lett.}\ }\textbf {\bibinfo {volume} {95}},\ \bibinfo {pages}
  {253004} (\bibinfo {year} {2005})},\ \Eprint
  {http://arxiv.org/abs/physics/0509106} {arXiv:physics/0509106
  [physics.atom-ph]} \BibitemShut {NoStop}%
\bibitem [{\citenamefont {Budker}\ \emph {et~al.}(2006)\citenamefont {Budker},
  \citenamefont {Lamoreaux}, \citenamefont {Sushkov},\ and\ \citenamefont
  {Sushkov}}]{PhysRevA.73.022107}%
  \BibitemOpen
  \bibfield  {author} {\bibinfo {author} {\bibfnamefont {D.}~\bibnamefont
  {Budker}}, \bibinfo {author} {\bibfnamefont {S.~K.}\ \bibnamefont
  {Lamoreaux}}, \bibinfo {author} {\bibfnamefont {A.~O.}\ \bibnamefont
  {Sushkov}}, \ and\ \bibinfo {author} {\bibfnamefont {O.~P.}\ \bibnamefont
  {Sushkov}},\ }\href {\doibase 10.1103/PhysRevA.73.022107} {\bibfield
  {journal} {\bibinfo  {journal} {Phys. Rev. A}\ }\textbf {\bibinfo {volume}
  {73}},\ \bibinfo {pages} {022107} (\bibinfo {year} {2006})},\ \Eprint
  {https://arxiv.org/abs/cond-mat/0511153} {arXiv:cond-mat/0511153 [cond-mat.other]}
\BibitemShut
  {NoStop}%
\bibitem [{\citenamefont {Eckel}\ \emph {et~al.}(2012)\citenamefont {Eckel},
  \citenamefont {Sushkov},\ and\ \citenamefont
  {Lamoreaux}}]{PhysRevLett.109.193003}%
  \BibitemOpen
  \bibfield  {author} {\bibinfo {author} {\bibfnamefont {S.}~\bibnamefont
  {Eckel}}, \bibinfo {author} {\bibfnamefont {A.~O.}\ \bibnamefont {Sushkov}},
  \ and\ \bibinfo {author} {\bibfnamefont {S.~K.}\ \bibnamefont {Lamoreaux}},\
  }\href {\doibase 10.1103/PhysRevLett.109.193003} {\bibfield  {journal}
  {\bibinfo  {journal} {Phys. Rev. Lett.}\ }\textbf {\bibinfo {volume} {109}},\
  \bibinfo {pages} {193003} (\bibinfo {year} {2012})},\ \Eprint
  {https://arxiv.org/abs/1208.4420} {arXiv:1208.4420 [physics.atom-ph]}
\BibitemShut {NoStop}%
\bibitem [{\citenamefont {Vutha}\ \emph {et~al.}(2018)\citenamefont {Vutha},
  \citenamefont {Horbatsch},\ and\ \citenamefont {Hessels}}]{Vutha:2018tsz}%
  \BibitemOpen
  \bibfield  {author} {\bibinfo {author} {\bibfnamefont {A.~C.}\ \bibnamefont
  {Vutha}}, \bibinfo {author} {\bibfnamefont {M.}~\bibnamefont {Horbatsch}}, \
  and\ \bibinfo {author} {\bibfnamefont {E.~A.}\ \bibnamefont {Hessels}},\
  }\href {\doibase 10.1103/PhysRevA.98.032513} {\bibfield  {journal} {\bibinfo
  {journal} {Phys. Rev. A}\ }\textbf {\bibinfo {volume} {98}},\ \bibinfo {pages}
  {032513} (\bibinfo {year} {2018})},\ \Eprint
  {http://arxiv.org/abs/1806.06774} {arXiv:1806.06774 [physics.atom-ph]}
  \BibitemShut {NoStop}%
\bibitem [{\citenamefont {Abel}\ \emph {et~al.}(2017)\citenamefont {Abel} \emph
  {et~al.}}]{Abel:2017rtm}%
  \BibitemOpen
  \bibfield  {author} {\bibinfo {author} {\bibfnamefont {C.}~\bibnamefont
  {Abel}} \emph {et~al.},\ }\href {\doibase 10.1103/PhysRevX.7.041034}
  {\bibfield  {journal} {\bibinfo  {journal} {Phys. Rev. X}\ }\textbf {\bibinfo
  {volume} {7}},\ \bibinfo {pages} {041034} (\bibinfo {year} {2017})},\
  \Eprint {http://arxiv.org/abs/1708.06367} {arXiv:1708.06367 [hep-ph]}
  \BibitemShut {NoStop}%
\bibitem{Graham2011axion}
P.~W.~Graham and S.~Rajendran, \href{\doibase 10.1103/PhysRevD.84.055013}{Phys.~Rev.~D~\textbf{84}, 055013 (2011)},\ 
\Eprint {https://arxiv.org/abs/1101.2691} {arXiv:1101.2691 [hep-ph]}. 
\bibitem{Stadnik2014axion}
Y.~V.~Stadnik and V.~V.~Flambaum, \href{\doibase 10.1103/PhysRevD.89.043522}{Phys.~Rev.~D~\textbf{89}, 043522 (2014)},\ 
\Eprint {https://arxiv.org/abs/1312.6667} {arXiv:1312.6667 [hep-ph]}. 
\end{thebibliography}%

\end{document}